\documentclass[12pt]{article}
\usepackage{epsfig}
\usepackage{graphics}
\usepackage{dcolumn}
\usepackage{amsmath}
\usepackage{cite}
\usepackage{changebar}
\usepackage[english]{babel}      
\usepackage{color}
\definecolor{green}{rgb}{0.00,1.00,0.00}
\definecolor{blue}{rgb}{0.00,0.00,1.00}
\definecolor{gray}{rgb}{0.75,0.75,0.75}
\definecolor{red}{rgb}{1.00,0.00,0.00}
\definecolor{darkgreen}{rgb}{0.00,0.39,0.00}
\definecolor{darkgrey}{rgb}{0.66,0.66,0.66}
\definecolor{darkblue}{rgb}{0.00,0.00,0.55}
\definecolor{darkred}{rgb}{0.55,0.00,0.00}

\newcommand{\be}{\begin{equation}}
\newcommand{\ee}{\end{equation}}
\newcommand{\ba}{\begin{eqnarray}}
\newcommand{\ea}{\end{eqnarray}}

\begin{document}
\renewcommand{\figurename}{{\bf Fig.}}
\renewcommand{\tablename}{{\bf Tab.}}
\sloppy
\title{\hfill {\normalsize\bf BI-TP 2003/29} \\ \vspace*{1cm}
Entanglement in Condensates involving Strong Interactions}
\author{David~E.~Miller$^{1,2}\,$\thanks{dmiller@physik.uni-bielefeld.de,
    \hspace*{2mm} om0@psu.edu}
 $\;$ and Abdel-Nasser~M.~Tawfik$^{1}\,$\thanks{tawfik@physik.uni-bielefeld.de} \\~\\
 {\small $^1$ Fakult\"at f\"ur Physik, Universit\"at Bielefeld, Postfach
 100131, D-33501
 Bielefeld, Germany} \\ 
 {\small $^2$ Department of Physics, Pennsylvania State University, 
         Hazleton, Pennsylvania 18201, USA} 
}
\date{}
\maketitle

\begin{abstract} 
We look at two well known examples of interacting systems relating to
condensed matter in which we put the strong interacting parameters. At high
quark chemical potentials and low temperatures we study the entropy arising
from the excitation in the BCS model of superconductivity and the Bose-Einstein
condensation (BEC) of colored quark pairs. We compare it 
with the ground state entropy for a system consisting of two colored
quarks. In the BCS model we found that 
the entropy strongly depends on the energy gap. Both
for the very small values of the momenta as well as those much  
greater than the characterizing Fermi momentum $p_f$, 
the ground state entropy is dominant. For the BEC case we 
suggest a phenomenological model to build up colored bosonic quark
pairs. Here the entropy entirely depends upon the short ranged repulsive
interactions between the quark pairs and vanishes for large momenta.   
\end{abstract}

\noindent
\begin{tabular}{rll}
PACS: & 12.39.-x & Phenomenological quark models \\
 & 03.75.Gg & Entanglement and decoherence in Bose-Einstein condensates \\
 & 03.75.Ss & Degenerate Fermi gases
\end{tabular}

\section{\label{sec:1} Introduction}

At high quark chemical potentials and low temperatures 
the hadronic matter has been conjectured to dissolve  
into degenerate fermionic quarks. 
The matter of very cold dense quarks might exist in the interior of 
compact stellar objects. Due to the difficulties of performing lattice
simulations with high chemical potentials, it is still not possible to
simulate the physics of these phases by using the lattice
QCD. Non-perturbative analysis at finite baryon density has been recently
carried out on the lattice by using the \hbox{Nambu-Jona-Lasinio}
model~\cite{HW}. The degenerate quarks near to the Fermi surfaces are 
generally expected to interact according to quantum chromodynamics so
that they can build up Cooper pairs. There are various mechanisms by which 
these quark pairs appear to be condensed, which depend on their total momenta. 
For example the Bardeen-Cooper-Schrieffer theory of superconductivity,
which is often referred to as just BCS~\cite{BCSa,BCSb}, can be applied
to the quarks in pairs with identical opposite momenta. Similarly, an 
extension of which is often called LOFF~\cite{LOFFa,LOFFb} is applied 
to the pairs of quarks with different momenta. Far away from the 
Fermi surfaces the quarks are blocked according to the Pauli principle. 
Therefore, they behave like free particles. In the present work 
we are interested only in the physics  
near to the Fermi surfaces where the quarks are attractively interacting. 
If we take into account the colors as the effective degrees of freedom 
while we keep all the other quantum numbers relating to simple symmetrical 
forms, the attractive interaction clearly leads to a 
breaking of the color gauge symmetry and therefore, the Cooper pairs get 
color superconducting. This structure are analogous to $SU(2)_c$ baryons which
are symmetric bosons of two colors. In the nuclear matter BCS pairings and
BEC have been studied some years ago~\cite{Roepke1,Roepke2}.  \\

     In this work we shall apply our previous calculations for the ground 
state entropy~\cite{Mill,MiTa1,MiTa2} on these states of quarks under such
extreme conditions. We shall compare the entropy 
arising from the excitations with that characterizing the ground state of
colored quark pairs. Here obviously we are dealing with a large
number of quark pairs. The other difference between this work
and~\cite{Mill,MiTa1,MiTa2} places in the nature of couplings between the
quarks. With the quantum entropy we mean the entropy arising from quantum
fluctuations. The latter, which differs from the thermal fluctuations,  
can also exists at zero temperature. Hereafter we refer to quantum entropy as
ground state entropy and {\it vice versa} whenever unambiguous. For the BCS
condensate we can directly apply the models given 
in~\cite{Mill,MiTa1}. But we simply consider the differences between
the nature of quark-pairs and their interactions at the Fermi level in BCS
and the structure and mixing of quarks in the colorless confined hadronic
singlet and octet states. For mixed states consisting of two colored
quarks, the quantum entropy is expected to be $T$ independent and equal $\ln
4$~\cite{Mill,MiTa1,MiTa2}. \\

Furthermore, since we have bosonic states consisting of quark pairs we can 
study a particular case of Bose-Einstein condensation. This condensate
entirely does not
depend upon the Fermi level. However, it represents a stimulating tool for the
understanding the extraneous condensates like BCS and LOFF, for
instance. Moreover, it is a rich phenomenological example since we know the
physics of BEC much better relative to the physics of the condensates at
such high quark chemical potential and very low temperature. Here we shall not
discuss the possible transition from color superconductivity of BCS at low density
to BEC of Cooper pairs at high density. In these two phases we individually 
calculate the entropy arising from the excitation for different momenta and 
compare it with that for the ground state. The transition from BCS to BEC 
has been discussed in many works~\cite{Roepke1,Roepke2,Itakura}.   

In the present work we shall use the usual statistical properties for 
the interacting many-particle Fermi and Bose gases with very short-ranged 
interactions~\cite{LaLi}. In this case one can utilize the second quantized
formalism to rewrite down the effects of the interaction by means of the canonical 
Bogoliubov transformation as a new gas with the same statistical properties
but with correspondingly modified energy spectrum.  \\

This article is organized as follows: The next section develops the model for
$SU(2)_c$ of symmetric bosons of two colors. In the two following sections
we present, respectively, the formulation for the Bose-Einstein
condensation and the BCS model of superconductivity. The next section is 
devoted to the presentation of the results followed by the discussion. 
Finally in the last section we state the conclusions.

\section{\label{sec:2}Model for $SU(2)_c$}

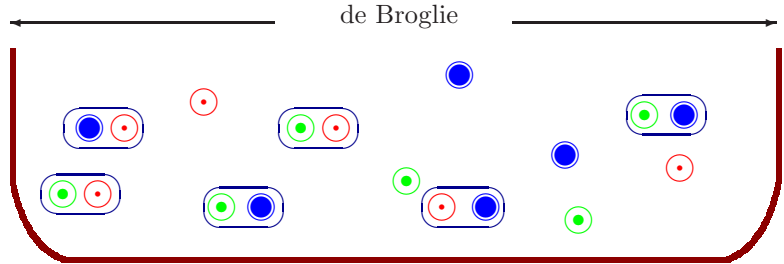
\begin{figure}
\begin{center}
\footnotesize
\begin{picture}(300,60)
{\color{green}{
\put(20,15){\circle{10}} 
\put(20,15){\circle*{4}} 
\put(110,40){\circle{10}} 
\put(110,40){\circle*{4}} 
\put(240,45){\circle{10}}
\put(240,45){\circle*{4}}
\put(80,10){\circle{10}} 
\put(80,10){\circle*{4}} 
\put(150,20){\circle{10}}
\put(150,20){\circle*{4}}
\put(215,5){\circle{10}}
\put(215,5){\circle*{4}}
}}
{\color{red}{
\put(30,15){\circle{10}} 
\put(30,15){\circle*{2}} 
\put(120,40){\circle{10}}
\put(120,40){\circle*{2}}
\put(160,10){\circle{10}} 
\put(160,10){\circle*{2}} 
\put(40,40){\circle{10}}
\put(40,40){\circle*{2}}
\put(250,25){\circle{10}}
\put(250,25){\circle*{2}}
\put(70,50){\circle{10}} 
\put(70,50){\circle*{2}} 
}}
{\color{darkblue}{
\put(20,15){\oval(30,15)}  
\put(110,40){\oval(30,15)} 
}}
{\color{blue}{
\put(85,10){\circle{10}} 
\put(85,10){\circle*{8}} 
\put(20,40){\circle{10}} 
\put(20,40){\circle*{8}} 
\put(245,45){\circle{10}}
\put(245,45){\circle*{8}}
\put(160,60){\circle{10}} 
\put(160,60){\circle*{8}} 
\put(200,30){\circle{10}} 
\put(200,30){\circle*{8}} 
\put(170,10){\circle{10}} 
\put(170,10){\circle*{8}} 
}}
{\color{darkblue}{
\put(158,10){\oval(31,15)}  
\put(22,40){\oval(30,15)}   
\put(235,45){\oval(30,15)}   
\put(75,10){\oval(30,15)}   
}}
\linethickness{1.7pt}
{\color{darkred}{ 
\qbezier(250,-10)(265,-5)(270,20)
\qbezier(0,-10)(-15,-5)(-20,20)
\linethickness{2.pt}
\put(0,-10){\line(250,0){250}}
\put(-19,20){\line(0,50){50}}
\put(271,20){\line(0,50){50}}
}}
\linethickness{1.pt}
\put(80,80){\vector(-1,0){100}}
\put(170,80){\vector(1,0){100}}
\put(105,80){de~Broglie}
\end{picture}
\end{center}
\caption{\sf Configuration of our phenomenological model for the Bose-Einstein
   condensation of coupling two colors as quark-pairs. Within the
   de~Broglie wavelength the enclosed system is colorless. The colored
   $q\;q$ pairs are coherently distributed. Different quark colors are
   represented by different solid circles and the pairs by ovals.  
\label{fig:1}
}
\end{figure}

The $SU(2)$ color structure has been investigated for finite baryon number
in a very special model involving the group characters~\cite{ElMiKa}. 
The thermodynamics of $SU(2)$ gauge theory with staggered fermions has been
studied at finite baryon density and zero temperature in the strong
coupling limit~\cite{DKM86}. A special property of the group structure of
$SU(2)$ is that the action remains real, meanwhile the basic group
structure for $SU(3)$ is complex. Analogous to $SU(2)_c$ symmetric colored
bosons we suggest a phenomenological model for the Bose-Einstein condensation in a
system consisting of tightly bound quark-pairs embedded within
nondegenerate fermions. We take the scale of the
order of the  de~Broglie wavelength $\lambda^2=h^2/(2\pi
m T)$. After having been accepted that the degenerate quarks build up
subsystems of atom-like 
pairs, we assume that the pairs are coherently distributed. Between each of
the two pairs there is a short-ranged repulsive interaction characterized by
the interaction strength $U_0$ which is usually given in units of energy
volume. Each quark 
is assumed to be coupled with its counterpart by a strongly attractive
force mediated by a kind of soft gluonic matter
that in nature might differ from the usual epoxy 
matter~\cite{Gery1} which supposed to hold the quarks in the confined hadronic
states. In the ground state of the system an arbitrary number of identical
quark pairs is allowed to occupy the same state. This homogenous
system in turn can be considered as a large number of individual and 
inseparable but still distinguishable subsystems, so that for binary 
interacting systems the canonical measure would be the entropy arising 
from the interaction. In other words for this bosonic subsystems we 
can estimate the entropy from the reduced density 
matrix as done in~\cite{vNeu,Mill,MiTa1}. This configuration might be  
illustrated as in Fig.~\ref{fig:1}. 
For the case where on this condensate an external potential is applied, the
system becomes inhomogeneous. Nevertheless, we are still able to consider it as
consisting of individual, inseparable and distinguishable subsystems. For
this model we suppose that all quark pairs are located within the de~Broglie
scale and the quarks exhibit coherent. Therefore, each of them can be
considered to be 
correlated with all others. And the correlations which are included 
into the quantum fluctuations are short as well as large ranged, so that  
they do not depend on the distance. For the simplicity we can only consider 
light quarks with binary correlations. Clearly the inclusion of
many-body interactions between the quark pairs leads to modification of
both ground and excited  
states and therefore their thermodynamics. In order to deal with this problem, 
we can apply the Hartree-Fock approximations in the second quantization 
formalism. The first is very familiar in the atomic physics, meanwhile 
the second is basically utilizing the many-body quantum field theory. 
The quark pairs are coupled with each other but their couplings 
in nature differ from those in the confined mesonic states. Nevertheless, 
their constituents are also supposed to be asymptotically free. 
Now we apply the Bose-Einstein statistics over these coupled states of two
colors and symmetric flavor.

\section{\label{sec:3}Formulation for Entanglement in \\Bose-Einstein condensation}

\begin{figure}
\centerline{\hspace*{-1cm}\includegraphics[width=16cm]{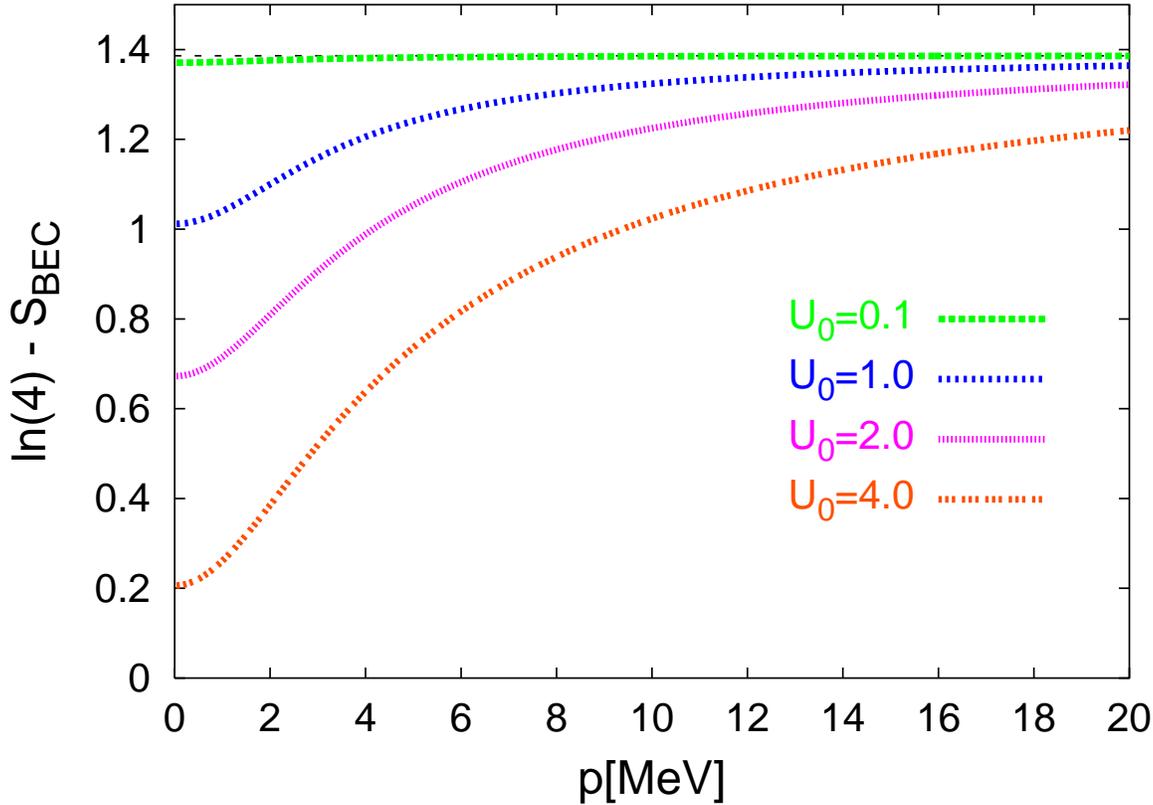}}%
 \caption{\label{fig:BEC}\sf The difference between the ground
   state entropy, $\ln 4$, and the entropy $S_{BEC}^{ent}$ of the excitation in
   BEC is given as a function of momentum $p$ and for different values of
   the effective interaction, $U_0$. For $p\rightarrow~0$,  
   $S_{BEC}^{ent}$ reaches its maximum value and it decreases with increasing
   $p$. For large $p$ the ground state entropy is dominant. For small $U_0$,
   $S_{BEC}^{ent}$ vanishes and the entropy is just given by the ground
   state value $\ln 4$.}
 \end{figure}

We look at the problem of entanglement in an interacting Bose gas
consisting of pairs of light quarks by using the second quantization 
formalism~\cite{Vedral}. The quarks effective degrees of freedom considered
in this work are just the colors. The flavors and other quantum numbers are
kept identical for both light quarks. For creation and annihilation operators,
$\hat{\Psi}^{\dagger}$, $\hat{\Psi}$ and effective interaction, $U_0$ the
Hamiltonian for interacting Bose gas can be written as
\ba
H &=& \int d{\mathbf r}\; \hat{\Psi}^{\dagger} \left[-\frac{\hbar^2}{2m} \nabla^2
  \hat{\Psi} + V\hat{\Psi}+\frac{U_0}{2} \hat{\Psi}^{\dagger} \hat{\Psi}
  \hat{\Psi} \right]  \label{bech1}\\
 &=& \sum_p \epsilon_p^0 a_p^{\dagger} a_p + \frac{U_0}{2V} \sum_{p_1,p_2, q}
 a_{p_1+q}^{\dagger} a_{{p_2}-q}^{\dagger}\; a_{p_2} a_{p_1} \label{bech2}
\ea
$\hat{\Psi}=\Psi+\delta\hat{\Psi}$ gives the quantum fluctuations whereas 
$\Psi$ is the single particle wavefunction.
The first term in Eq.~\ref{bech1} represents the non-interacting part of the
Hamiltonian. In Eq.~\ref{bech2} the Hamiltonian of the degenerate and nearly
ideal Bose gas 
the operators are given in momentum space.  
Simultaneously, we have inserted two bosonic 
creation and annihilation operators, $a_p^{\dagger}$ and  $a_p$,
respectively for the 
usual spatially dependent operators in the previous equation. 
$U_0$ is the interaction strength from the collision of two particles with
momenta, $p_1$ and $p_2$, which produce two new particles with momenta,
$p_{1}+q$ and $p_2-q$, respectively. Here we have assumed that the interaction
does not depend on the momentum change  $q$~\cite{PS}.   
With the Bogoliubov canonical transformations we get a Hamiltonian of
non-interacting system. The latter is solvable and therefore we can estimate its
eigenvalues. The second quantized operators in $p$ direction, $a,
a^{\dagger}$ and in $p$ direction, $b, b^{\dagger}$ are  commutative, so
that we can replace $a$ with $b$ in $H$, which remains  
invariant. In doing the Bogoliubov transformations, we introduce two variables,
$\alpha$ and $\beta$, where  
\ba
\alpha &=& u(p) a + v(p) b^{\dagger} \\
\beta  &=& u(p) b + v(p) a^{\dagger} 
\ea
and the $p$-dependent variables $u(p)$ and $v(p)$ are just the
coefficients, which will be defined in Eq.~\ref{BogU},~\ref{BogV}.
$\alpha$ and $\beta$, and their counterparts in $-p$ direction are the
operators, which create and destroy the elementary excitations
(quasi-particles). We choose the phases of real $u(p)$ and $v(p)$ so that the
interacting  parts in $H$ are entirely removed. In doing this, we
diagonalize $H$ and therefore we can set
\hbox{$a_p=u(p)\alpha-v(p)\alpha_{-p}^{\dagger}$} and \hbox{$a_{-p}=u(p)\alpha_{-p}-v(p)\alpha_p^{\dagger}$}. 
\ba
H=\frac{N^2 U_0}{2 V} + \sum_{p\neq 0} \epsilon_p^0 \alpha_p^{\dagger}
\alpha_p - \frac{1}{2} \sum_{p\neq 0} \left(\epsilon_p^0 - n_0 U_0
  -\epsilon(p)\right) 
\ea
$\epsilon_p^0$ is the single-particle energy and $\epsilon(p)$ will be given in
Eq.~\ref{esplp}.   
As given in~\cite{Vedral} the ground state in such entangled collection of
decoupled Boson pairs in $p$ and $-p$ directions is given as
\ba
|\psi_0>&=& g \prod_{p\neq 0} \frac{1}{u(p)} \sum_{i=0}^{\infty}
\left(-\frac{v(p)}{u(p)}\right)^i \; |n_{-p}=i; n_{p}=i>
\ea
and the entropy from the excitation of two bosons with equal
and oppositely directed momenta $p$ is given by
\ba
S_{BEC}^{ent} &=& g \sum_{p\neq0} \left\{
  \frac{\ln\left(\frac{u(p)}{v(p)}\right)^2}{\left(\frac{u(p)}{v(p)}\right)^2-1}
  \;\;-\;\;\ln\left(1-\left(\frac{v(p)}{u(p)}\right)^2 \right)\right\}
\label{eqBEC1}
\ea
where $g$ is the degeneracy factor and the two coefficients of the Bogoliubov
canonical transformation are defined as  
\ba
u(p) &=& \left[\frac{1}{2}
  \left(\frac{\zeta(p)}{\epsilon(p)}+1\right)\right]^{1/2} \label{BogU}\\ 
v(p) &=& \left[\frac{1}{2}
  \left(\frac{\zeta(p)}{\epsilon(p)}-1\right)\right]^{1/2} \label{BogV}   
\ea
From Eq.~\ref{BogU}, \ref{BogV}, we get \hbox{$u(p)^2-v(p)^2=1$} which
  assures that the energy over the space of 
the amplitudes for unoccupation $u$ and occupation $v$ is minimum. 

Alternatively, we can obtain these results, if we directly diagonalize the 
Hartree-Fock Hamiltonian together with the self-consistency equation of the
order parameter $\Delta$. For the interacting
Bose gas of quark-pairs with particle density $n=N_0/V_0>0$, we apply the
following definitions: 
\ba
\zeta(p)    &=&\epsilon_0(p)+nU_0 \label{zeeta}\\ 
\epsilon(p) &=& \left(\epsilon_0(p)^2+2\epsilon_0(p) n U_0\right)^{1/2}
             \label{esplp} 
\ea
\hbox{$\epsilon_0(p)=(p^2+m^2)^{1/2}$} - as given above - is the energy of
single-particle excitation and $m$ is the reduced mass of the
quark-pair. The validity of above definitions in Eq.~\ref{zeeta},
~\ref{esplp} is guaranteed  by \hbox{$nU_0 \geq \epsilon_0(p)$}. After
plugging into Eq.~\ref{eqBEC1} we find that 
\ba
S_{BEC}^{ent}(p) &=& g \;\sum_{p\neq0}\left\{
      \frac{\zeta(p)-\epsilon(q)}{2\,\epsilon(p)}\;
      \ln\left(\frac{\zeta(p)+\epsilon(p)}{\zeta(p)-\epsilon(p)}\right)
      \right. \nonumber \\
  & &\hspace*{32mm} +\left.\ln\left(\frac{\zeta(p)+\epsilon(p)}{2\,\epsilon(p)}\right) \right\}
\label{eqBEC2}
\ea 
the sum of the excitation entropy for each boson pair with the momentum
$p$. \\ 

\subsection{Ground state entropy of colored quark-pairs}

In the limit of low temperature $T$ and small distances $R$ between the two
static quarks, the free energy for the $q\;q$ system is presently being
studied in $SU(3)$ pure gauge theory on the lattice~\cite{Vogt1}.     
By taking the colors $N_c$ as the effective degrees of freedom, the
quantum entropy in the ground state of a system consisting of two colored
quarks is given as~\cite{Mill,MiTa1,MiTa2} 
\ba
S_{qq} &=& \ln N_c^2 = \ln 4\label{eq:ln4}
\ea
This value is clearly temperature independent. Moreover, we should
mention that for $T=[0,m]$ it is all-dominant against the thermal
entropy and as discussed in~\cite{Mill,MiTa1,MiTa2}, for this reason it is 
used to be abstracted away for high temperatures. But if we look at a
system of massive quarks~\cite{SZ,BLRS}, the
region of temperatures where the ground state entropy remains significant will be
correspondingly large. The effects of the quark mass on the color
superconductivity~\cite{dBlaschke1} and on the stability of hybrid
stars~\cite{Igor1} are recently reported. Thus we think that the
ground state entropy is an essential physical observation in understanding
the compact steller objects and the coled quark matter with very high chemical
potential.     

To compare this constant value with the entropy arising from the excitation
in the BEC condensate, we define 
\ba
S_{BEC}(p) &=& S_{qq} - S_{BEC}^{ent}(p) = \ln 4 -
                        S_{BEC}^{ent}(p)\label{eq:ln4minus} 
\ea
The results
are given in Fig.~\ref{fig:BEC}. $S_{BEC}^{ent}(p)$ is maximum when
$p\rightarrow 0$, i.e. $m >> p$ in $\epsilon_p^0$ and correspondingly,
$u(p)\sim v(p)$. For high momenta, $S_{BEC}^{ent}(p)\rightarrow 0$, and
$u(p)\rightarrow 1$ meanwhile $v(p)\rightarrow 0$. But as we will see,
this behavior is further strongly depending on the interaction strength, $U_0$.   

For \hbox{$U_0 \rightarrow 0$}, the entropy \hbox{$S_{BEC}^{ent}(p)\rightarrow
  0$}. Thus for constant momentum the controlling parameter over
  $S_{BEC}^{ent}$ is the $U_0$ which is defined as the interaction strength  
in the units of energy volume. For the noninteracting  BEC of an ideal gas
consisting the coupled quark pairs \hbox{$S_{BEC}^{ent}=0$}. Clearly in this
case the excitation entropy, $S_{BEC}^{ent}$, is a phenomenon which does not 
depend on the Fermi surface.

\section{\label{sec:4}Formulation for Entanglement \\in the BCS Model}

\begin{figure}
\centerline{\hspace*{-1cm}\includegraphics[width=16cm]{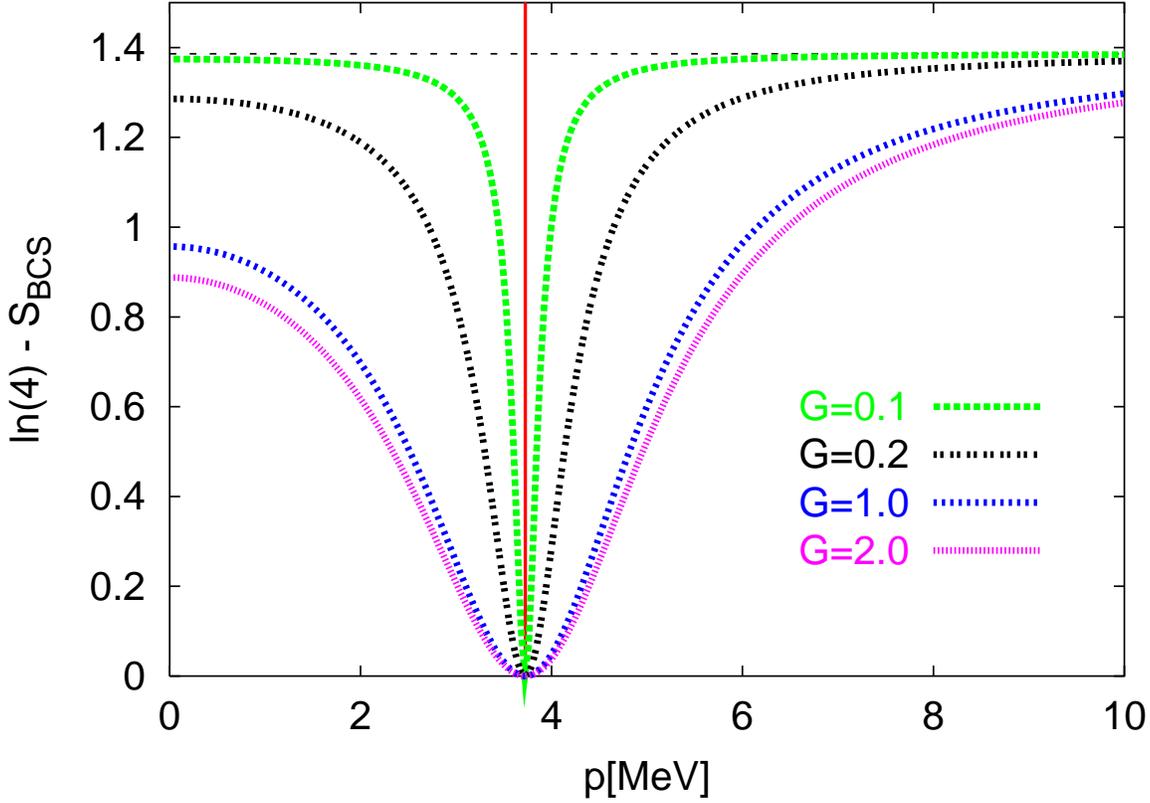}}%
 \caption{\label{fig:BCS}\sf The difference between $\ln 4$ (quantum
   entropy for a system of two colored quarks) and the entropy
   $S_{BCS}^{ent}$ which arrises from the excitation is given for different
   momenta $p$ and couplings $G$. We note that $S_{BCS}^{ent}$ strongly
   depends on the  momentum space and the energy gap. At the Fermi level
   $p_f$ (vertical line), $S_{BCS}^{ent}$ reaches a maximum value, $\ln 4$.
   This is valid for all couplings and the 
   correspondingly energy gaps. Obviously, for small couplings it vanishes
   both below as well above $p_f$.   
   }
 \end{figure}

As introduced in section~\ref{sec:1} the overlap between BCS theory and 
the phenomenon of BEC will not be taken directly into consideration 
in this work. For the entropy in the paired phase 
we look at the ground state entropy which arises from the
correlations in order to determine the full extent of the
entanglement of the ground state. The expectation is that the extent of 
the entanglement is determined by the mixing as we have determined
in~\cite{Mill,MiTa1,MiTa2} and the excitations in the ground state 
of BCS, which analogously arises from the electron-phonon interaction.
As a theoretical replacement for this coupling we have the octet structure 
which couples through the gluons to the particular colors. This is true 
for all the gluons building colored pairs except for the coupling generated 
from the Gell-Mann matrix $\lambda_8$. In the second quantized formalism the
interactions can be related to the entanglement~\cite{Vedral}. We use the
description of superconductors~\cite{Fey,PS}, which can be treated in a
similar way to the interaction causing a 
Bose-Einstein condensation of the Cooper pairs. 
For finite couplings nonzero bound state energies exist. 
There is another basic feature of BCS which involves oppositely directed
momenta and spins above and below the Fermi level.
As we have discussed in the previous section,
we carry out the Bogoliubov canonical transformation for the field 
amplitudes deoccupation and occupation of the states, respectively, 
\ba
u(p) &=& \left[\frac{1}{2}\; \left(1+
    \frac{\zeta(p)}{\epsilon(p)}\right)\right]^{1/2}, \\ 
v(p) &=& \left[\frac{1}{2}\; \left(1-
    \frac{\zeta(p)}{\epsilon(p)}\right)\right]^{1/2},  
\ea 
and therefore, \hbox{$u(p)^2+v(p)^2=1$}. We find by 
taking into account the proper statistics 
\ba
S_{BCS}^{ent}(p) &=& g \;\sum_{p\neq0}\left\{
  \frac{\ln\left(\frac{u(p)}{v(p)}\right)^2}{\left(\frac{u(p)}{v(p)}\right)^2+1}
  + \ln\left(1+\left(\frac{v(p)}{u(p)}\right)^2 \right)\right\} \label{eqBCS1}
\ea
where
\ba
\zeta(p) &=& \epsilon_0(p)-\mu, \\
\epsilon(p) &=& \left(\zeta(p)^2+\Delta^2\right)^{1/2}
\ea 
and $\mu$ is the quark chemical potential.  We call to mind that the change
of the minus signs 
in Eq.~\ref{eqBEC1} to the plus signs here arises on account of the Fermi
statistics in the present case. 

This dispersion relation is to be modified by the existence of coherent
effects (pairings) at $T=0$. $\epsilon(p)$ is equivalent to the Bogoliubov  
quasiparticle energy, which according to $\mu$ characterizes the minimum of the
energy gap. $\Delta$ is an interval in which no eigenenergies are allowed
in the one-particle energy spectrum. It is usually called the energy gap
and plays the role of order parameter. 
\ba
S_{BCS}^{ent}(p) &=& g \sum_{p\neq0} 
    \left\{\frac{\epsilon(p)-\zeta(p)}{2\;\epsilon(p)} 
    \ln\left(\frac{\epsilon(p)+\zeta(p)}{\epsilon(p)-\zeta(p)}\right)\right.
    \nonumber \\   
 & & \hspace*{29mm}-\left.\ln\left(\frac{\epsilon(p)+\zeta(p)}{2 \epsilon(p)}\right) 
\right\}
\label{eqBCS3}
\ea

\noindent
In Eq.~\ref{eqBEC2}, \ref{eqBCS3} we can replace $\sum_{p\ne0}$ by
$V/(2\pi^2) \int p^2 \, dp$ and set the degeneracy factor
\hbox{$g=2$}. 

Eq.~\ref{eqBCS3} gives the sum of the entanglement entropy of BCS pairs
with momenta $]0,p]$. In order to evaluate this entropy arsing from the
entanglement in the ground state, we must know more about the system. In
the BCS limit, we set 
the chemical potential \hbox{$\mu\equiv\epsilon_f=(\pi^2\,  n^{2/3} +
  m^2)^{1/2}$}, the Fermi level at vanishing temperature. The gap in the
spectrum of single-particle excitation which controls the sign of chemical
potential, $\mu$,  can be found by solving the following integral
equation~\cite{PS,Rischke1}:  
\ba
1 &=& \frac{3 U_0 V}{2 \pi^2} \int_0^{\infty} p^2 dp
\left[\frac{\epsilon(p)-\zeta(p)}{\epsilon(p) \zeta(p)}\right].
\label{eqBCS4}
\ea 
We note that $\zeta(p)$ is the kinetic energy from the Fermi level and
clearly depends upon the chemical potential. Therefore, for the excitation
energy in the ground state $\Delta$ can be parameterized as follows: 
\ba
\Delta &=& \epsilon_f \;\left(\frac{2}{e}\right)^{7/3}\; 
\exp\left(\frac{-1}{4\;G}\right) 
\label{eqBCS5}
\ea  
where $\epsilon_f$ is the Fermi energy, which per definition at $T=0$
is equivalent to the ground state quark chemical potential
$\mu_f$. $G$ is the coupling strength, which represents the
controlling parameter in the BCS model. For vanishing $G$ we have
\hbox{$\Delta=0$} and then \hbox{$S_{BCS}^{ent}=0$}. As in
Eq.~\ref{eq:ln4minus} we define 
\ba
S_{BCS}(p) &=& \ln 4 - S_{BCS}^{ent}(p) \label{eq:ln4bcsminus}
\ea
This entropy difference is given in Fig.~\ref{fig:BCS} for different
couplings and momenta. 
So far we can conclude that the part of the entanglement entropy is entirely
arising from the interactions/excitations between the quark pairs with
opposite momenta at the surface of Fermi sea. We can also conclude that at
the Fermi surface the maximum excitation entropy equals to the entropy for
the mixing of two colored quarks (quantum entropy). The latter is $T$ and
$p$ independent and all-dominant for $T=[0,m]$.

\section{\label{sec:5}Results and discussion}

In Fig.~\ref{fig:BEC} the difference between the entropy of the excitation in 
BEC of two colored bosonic quark-pairs (Eq.~\ref{eqBEC2}),
which are characterized by the phenomenological model given in
section~\ref{sec:2}, and the ground state entropy (Eq.~\ref{eq:ln4}) is
depicted in dependence upon the momenta $p$ and for different interaction
lengths $U_0$. Here Eq.~\ref{eqBEC2} is numerically solved for equal
successive momentum intervals $p$. We note that $S_{BEC}^{ent}$ starts from
a maximum value at $p\rightarrow 0$ and decays with increasing $p$. For
large $p$ it radically vanishes indicating no contribution to the entropy
from the excitation, since for large momentum the scaling exceeds the
de~Broglie wavelength and therefore the correlation between the
bosonic quark pairs entirely disappears.  
We also notice how the entropy difference depends on the interaction strength
$U_0$. Larger $U_0$, smaller is the difference at $p\rightarrow 0$ and slower is
the increasing towards the asymptotic value, $\ln 4$. For $U_0\rightarrow 0$, we
have from Eq.~\ref{eqBEC2} that $S_{BEC}^{ent}\rightarrow 0$. 
Thus we can conclude that the interaction strength $U_0$ fully determines
the properties of the $S_{BEC}^{ent}$ as the entropy arising from the
entanglement. Therefore, it can be taken as the controlling parameter. 
On the other hand, the asymptotically vanishing value of $S_{BEC}^{ent}$ 
in the condensate of quark pairs disappears for large momenta since
there is no remaining interaction between the quark pairs.  \\

The BCS calculations are given in Fig.~\ref{fig:BCS}. The difference
between the ground state entropy that arises from the
entanglement in the BCS (Eq.~\ref{eqBCS3}) is depicted as 
a function  of momenta $p$ for different values of the coupling $G$. 
We should notice again that $S_{BCS}^{ent}$ is highly structured. 
There are  two regions where $S_{BCS}^{ent}$ vanishes: 
one is well below while the other is well above the Fermi momentum $p_f$,
which is shown as the entropy difference from the ground state value $\ln 4$. 
The peak, which is located around the Fermi surface, is shown by the sharp dip.
When the value for $S_{BCS}^{ent}$ starts to move above its vanishing
value for the momenta higher in the Fermi sea. 
Near the Fermi surface the excitation entropy rapidly increases. 
This behavior reflects the appearance of correlations between the quarks below
and their counterparts above $p_f$. The asymptotic value is again just $\ln 4$, 
the value of the quantum entropy for the mixing in a system of two
colored quarks (Eq.~\ref{eq:ln4}). As discussed  above this behavior 
strongly depends on the coupling $G$ and thereby on the gap parameter
 $\Delta$. Nevertheless, the general structure remains much the same over
many values. For larger $\Delta$, the values of $S_{BCS}^{ent}$ get larger
even at quite small and rather large $p$. Thus the asymptotic region will
be moved towards much larger momenta. Nevertheless, the maximum value 
of $S_{BCS}^{ent}$ at $p_f$ does not itself depend on $G$. 
Therefore, this value characterizes the excitation at the Fermi surface. 
The entropy of excitation at $p_f$ is the same as the value $\ln 4$ of the 
mixing of the paired colored quarks at $T=0$, which is the size of the dip. 
Therefore, we can accordingly conclude that the condensate of BCS
pairings can be a measure of the density of pair states around the Fermi
surfaces. Thus in this model we would expect that as soon the gluonic 
interactions reach the Debye cutoff momenta that then $S_{BCS}^{ent}$ 
also vanishes as does the excitation entropy. 

The quark pairs in this BCS model are characterized by asymmetric color 
degrees of freedom and equal opposite momenta and spins. 
All the other degrees of freedom are kept identical. 
One quark is located just below the Fermi level,
while its counterpart is found just above it. 
The coupling $G$ and therefore the
energy gap $\Delta$ are likewise model dependent~\cite{Bowers1}. 
Hence we are left with parameterizing these quantities. 
Eq.~\ref{eqBCS5} gives the parameterization of
$\Delta$ in relation to $G$ for symmetric flavors, momenta and spins and
asymmetric colors. This relationship signifies that for vanishing $G$,
\hbox{$\Delta\rightarrow 0$}. With increasing $G$ the energy gap $\Delta$
increases too. $\Delta$ can be illustrated as region around the Fermi
level, in which $S_{BCS}^{ent}$ is finite. Only those quarks whose momenta
fit into this region are considered. In other words large $\Delta$ leads
to large correlations with correspondingly large excitations between the
quark pairs around $p_f$. Therefore, there are more pairs 
which are contributing to the value of $S_{BCS}^{ent}$. Vanishing values
of $\Delta$, on the other hand, 
lead to an absence of interacting quarks at the Fermi surface. 
Hence, the theoretical system of deconfined {\it free} quarks 
at these high quark chemical potentials and low temperatures 
can properly be considered as a closed pure state without any mixing
structure~\cite{Mill,MiTa1,MiTa2}. This kind of system has according 
to Nernst's heat theorem zero entropy in the low temperature limit. 

Another worthwhile result appearing in this figure is the shift of the 
asymptotically zero value of $S_{BCS}^{ent}$ towards higher momenta
for larger $G$. This situation happens because the larger values of $\Delta$ 
lead to inclusion a larger number of quarks deep within the
Fermi surfaces and simultaneously more quarks with momenta greater than
$p_f$. Then for much larger $\Delta$ we can freely move to higher momenta.
Therefore, we notice for these cases that the region of nonzero
values of the excitation entropy $S_{BCS}^{ent}$ becomes correspondingly wider.     
  
\section{\label{sec:6}Conclusion}

An important consideration in relation to this work is the existence of the
excitations at $T=0$. The only quark interactions allowed at $T=0$ are those
of degenerate fermions near the Fermi surface. Besides this we should
consider the mixing of colored quarks. A purely formal
clarification explains the 
excitations as a result of the Bogoliubov canonical transformation taking
the interacting Hamiltonian with pairs 
of oppositely directly momenta for the ground state operators into a simple
sum over all finite momenta of the quasiparticle number operators with a 
modified energy spectrum containing various pieces of the interactions. These
excitations arise in color space in much the same way that the spin waves appear 
in the similar operators for atomic physics. Clearly these excitations are not
so simply represented in the $SU(3)_c$ space in the actual colorless ground 
state of the hadrons, which has a singlet representation. However, the octet
states are mostly those involving only two colors in the Gell-Mann matrices
-- except for ${\lambda}_8$. Thus by taking linear combinations of ${\lambda}_3$ 
and ${\lambda}_8$ we may write down {\it nine} matrices for three pairs of $SU(2)$
color pair states which have been constructed from the octet states. Each pair
may be treated analogously to the spin waves. All three together could be
imagined to be waves propagating in three perpendicular directions, which are
clearly not independent of each other. However, over very short distances
a large range of momenta could be accommodated, which within these values of 
momenta the oppositely directed pair of constituents are present.

Finally, we conclude by noting that the effects of the interparticle
interaction for both the Bose-Einstein condensate and the BCS superconductor
causes the entanglement. This fact is characterized by the finite momentum
contributions to the condensation process which offer the Bogoliubov canonical
transformation results in the excitation spectra. It is the coefficients
of this transformation that appear in the quantum entropy density of entanglement.
Thus we have noted that for both the composite Bose-Einstein condensate
and the superconductor with its Cooper pairing structure the entanglement
arises from the interaction between the oppositely directed momenta of
the constituent particles. For the parameters we used here the maximum entropy for
the excitations in the two condensates is compatible with the value for the
mixing of two quarks at zero temperature. Therefore, the entropy arising
from the excitations in the condensates BCS and BEC, $S\leq\ln 4$, depend
upon the momentum space.\\ \\

{\bf Acknowledgments}\\
In preparing this work we have benefited from the stimulating discussions with
Krzysztof~Redlich. We acknowledge the further helpful discussions with 
David~Blaschke, Robert~Pisarski, Dirk~Rischke, Helmut~Satz and Igor~Shovkovy.
D.E.M is very grateful to the support from the
Pennsylvania State University Hazleton for the sabbatical leave of absence
and to the Fakult\"at f\"ur Physik der Universit\"at
Bielefeld, especially to Frithjof Karsch.

\end{document}